\documentclass[aip, apl, 
reprint,
amsmath,amssymb]{revtex4-1}

\usepackage{graphicx}
\usepackage{dcolumn}
\usepackage{bm}

\usepackage{epstopdf} 

\begin{document}

\preprint{\today}
\title{Coherent Heteroepitaxy of Bi$_2$Se$_3$ on GaAs (111)B}

\author{A. Richardella}
\author{D. M. Zhang}
\author{J. S. Lee}
\author{A. Koser}
\author{D. W. Rench}
\affiliation{Department of Physics, The Pennsylvania State University, University Park, PA 16802, USA}
\author{A. L. Yeats}
\author{B. B. Buckley}
\author{D. D. Awschalom}
\affiliation{Department of Physics, University of California, Santa Barbara, CA 93106 USA}
\author{N. Samarth}
\email{nsamarth@psu.edu}
\affiliation{Department of Physics, The Pennsylvania State University, University Park, PA 16802, USA}

\date{\today}

\begin{abstract}
We report the heteroepitaxy of single crystal thin films of Bi$_2$Se$_3$ on the (111)B surface of GaAs by molecular beam epitaxy. We find that Bi$_2$Se$_3$ grows highly c-axis oriented, with an atomically sharp interface with the GaAs substrate. By optimizing the growth of a very thin GaAs buffer layer before growing the Bi$_2$Se$_3$, we demonstrate the growth of thin films with atomically flat terraces over hundreds of nanometers. Initial time-resolved Kerr rotation measurements herald opportunities for probing coherent spin dynamics at the interface between a candidate topological insulator and a large class of GaAs-based heterostructures.
\end{abstract}


\maketitle

The narrow band gap semiconductor Bi$_2$Se$_3$ has recently emerged as a promising basis for creating a state of matter known as a topological insulator (TI) wherein protected states can be produced at the surface of the material via the locking of spin and momentum by the constraints of time reversal symmetry.\cite{Fu:2007fk,Qi2008,Zhang2009a} The prediction that it has the requisite electronic structure for forming these special conducting surface states spanning its bulk electronic energy gap has been confirmed by angle resolved photoemission spectroscopy.\cite{Zhang2009a, Xia2009,Hsieh2009} With a bulk band gap ($\sim$ 0.3 eV) larger than other relevant materials, Bi$_2$Se$_3$ is one of the best candidate materials for engineering of the Fermi energy into the bulk band gap so that transport can occur only through these surface states. However, this simple prescription has proved hard to realize because of an inherent tendency of the material to form Se vacancies or antisites that serve as donors,\cite{Navratil2004} moving the Fermi energy far above the gap and making the contribution of the surface states to transport properties difficult to detect.\cite{Hor2009} 

The growth of Bi$_2$Se$_3$ by molecular beam epitaxy (MBE) provides a potentially attractive solution for minimizing such defects by allowing for flexible control of growth conditions. To date, MBE growth of Bi$_2$Se$_3$ has been demonstrated on several substrates, including silicon, graphene and SrTiO$_3$, albeit without complete removal of midgap states.\cite{Zhang2009, Li2010, Zhang2010, Song2010, Chen2010} For silicon, the MBE growth of single crystal Bi$_2$Se$_3$ requires the introduction of an intermediate layer (e.g. a monolayer of Bi or amorphous layers) that improves the film quality by effectively decoupling it from the substrate, while graphene is conductive, complicating transport measurements of the surface states. In this Letter, we report the heteroepitaxy of Bi$_2$Se$_3$ thin films upon another technologically important substrate material, GaAs. Notably, we show that the epitaxial growth is coherent with the substrate, thus opening routes for exploring the coupling of spin polarized TI states with electronic states in a wide variety of advanced semiconductor heterostructures, including magnetically doped III-V and II-VI semiconductors. 

We carried out MBE growth of Bi$_2$Se$_3$ thin films on epiready, semi-insulating GaAs (111)B substrates using thermal evaporation of high purity (5N) elemental Bi and Se from conventional Knudsen cells. After thermal desorption of the native oxide on the substrate under an arsenic flux, we first deposited a very thin GaAs buffer layer ($\sim$18 monolayers), yielding a very flat GaAs surface without the pitting of the surface that occurs with desorption of the oxide or the three dimensional hillocks that form with thicker buffers.\cite{Sugahara2003}  Bi$_2$Se$_3$ was then grown at a substrate thermocouple temperature of 400 $^\circ$C (corresponding to an estimated actual substrate temperature of $\sim 320 ^\circ$C)  and a Se:Bi beam equivalent pressure ratio ranging from $\sim$ 10:1 to $\sim$ 30:1. 

\begin{figure}[]
\includegraphics[width=50mm]{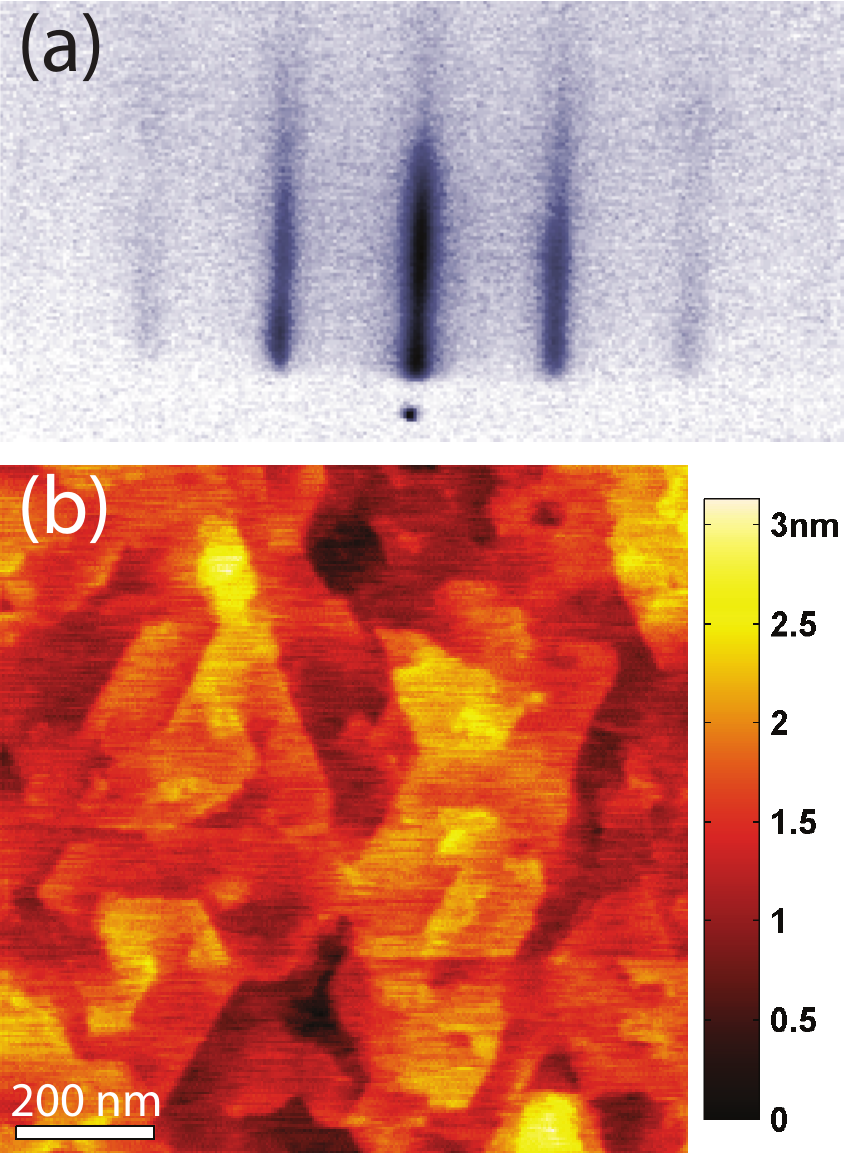}
\caption{\label{AFM}(Color online)(a) Streaky, unreconstructed RHEED diffraction patterns such as this were usually observed along principal crystalline directions during and after the film growth. (b) AFM image of the surface of a Bi$_2$Se$_3$ film. Large terraces hundreds of nm across can be seen whose $\sim 1$ nm step heights are consistent with single QLs.  The RMS roughness of the film is $\sim 0.5 $nm. }
\end{figure}

Bi$_2$Se$_3$ has a tetradymite, trigonal crystal structure with a rhombohedral unit cell that can be viewed as consisting of three sets of groupings of Se-Bi-Se-Bi-Se planes commonly referred to as quintuple layers (QLs). Each Se or Bi plane within the QL is a two dimensional hexagonal lattice. This matches the hexagon structure of the GaAs (111) surface with a lattice mismatch of 3.55\%. We have grown Bi$_2$Se$_3$ films ranging in thickness from $\sim 30$ nm down to a few QLs ($\sim 3$ nm) with a typical growth rate of $\sim$0.85 QL/min. Reflection high energy electron diffraction (RHEED) measurements during growth of Bi$_2$Se$_3$ indicate an unreconstructed surface (Fig.~\ref{AFM}(a)). We have also observed RHEED oscillations of the specular spot (data not shown), with each oscillation corresponding to the growth of a QL, indicating that the Bi$_2$Se$_3$ thin films grow layer-by-layer.\cite{Song2010, Li2010} 

\begin{figure}[]
\includegraphics[width=50mm]{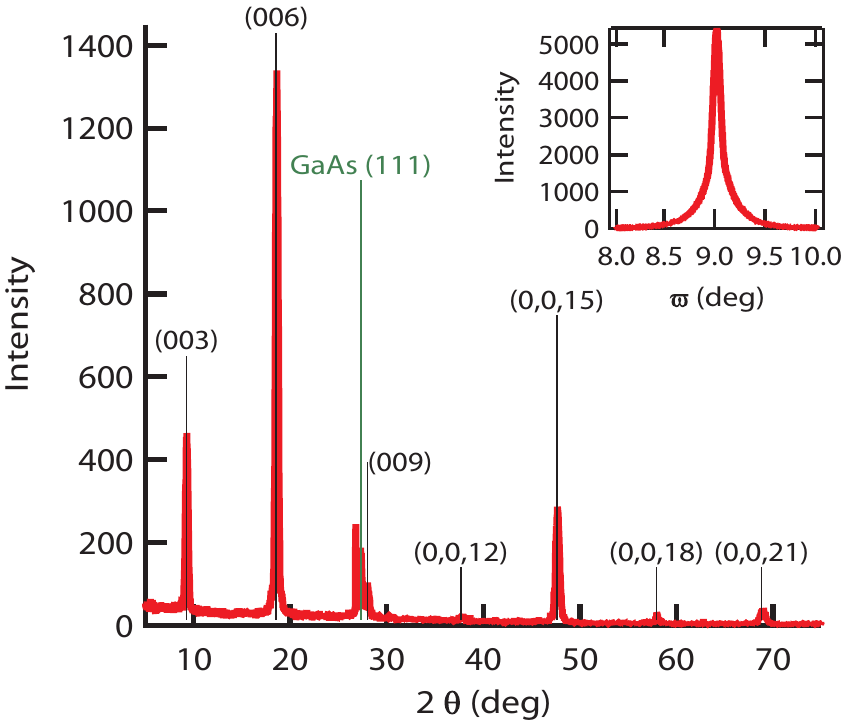}
\caption{\label{XRD}(Color online) X-ray diffraction of a $\sim$25 nm thick Bi$_2$Se$_3$ film. The (003) family of reflections shows that the films are highly c-axis oriented. Bi$_2$Se$_3$ peaks are labeled from ICDD PDF file 00-033-0214. Inset shows the rocking curve of the (006) reflection giving a FWHM of $ 0.1^{\circ}$.}
\end{figure}

The morphology of the films was studied \textit{ex-situ} by atomic force microscopy (AFM). For some films, like the 25 nm thick film shown in Fig.~\ref{AFM}(b), we grew a second buffer of ZnSe, only a few monolayers thick, by atomic layer epitaxy. While we were unable to directly confirm the presence of ZnSe in these samples by x-ray diffraction (XRD) or Raman spectroscopy, they did tend to result in very flat Bi$_2$Se$_3$ surfaces with RMS roughnesses of $\sim 0.5$ nm. Samples without the ZnSe buffer were slightly rougher with an average RMS roughness of a few nm. Very thin films of 2-3 QLs appear to exhibit island-like growth, similar to observations made for growth of Bi$_2$Se$_3$ on graphene.\cite{Song2010} XRD measurements show reflections only from the (003) family of planes of the film, indicating that the films are highly c-axis oriented along the growth direction (Fig.~\ref{XRD}). The rocking curve yielded a full width half maximum of 0.1$^{\circ}$, significantly better than those reported for growth on vicinal Si substrates with an amorphous layer.\cite{Li2010}. While including a ZnSe buffer resulted in a flatter film, it also resulted in a wider rocking curve. 

\begin{figure}[]
\includegraphics[width=40mm]{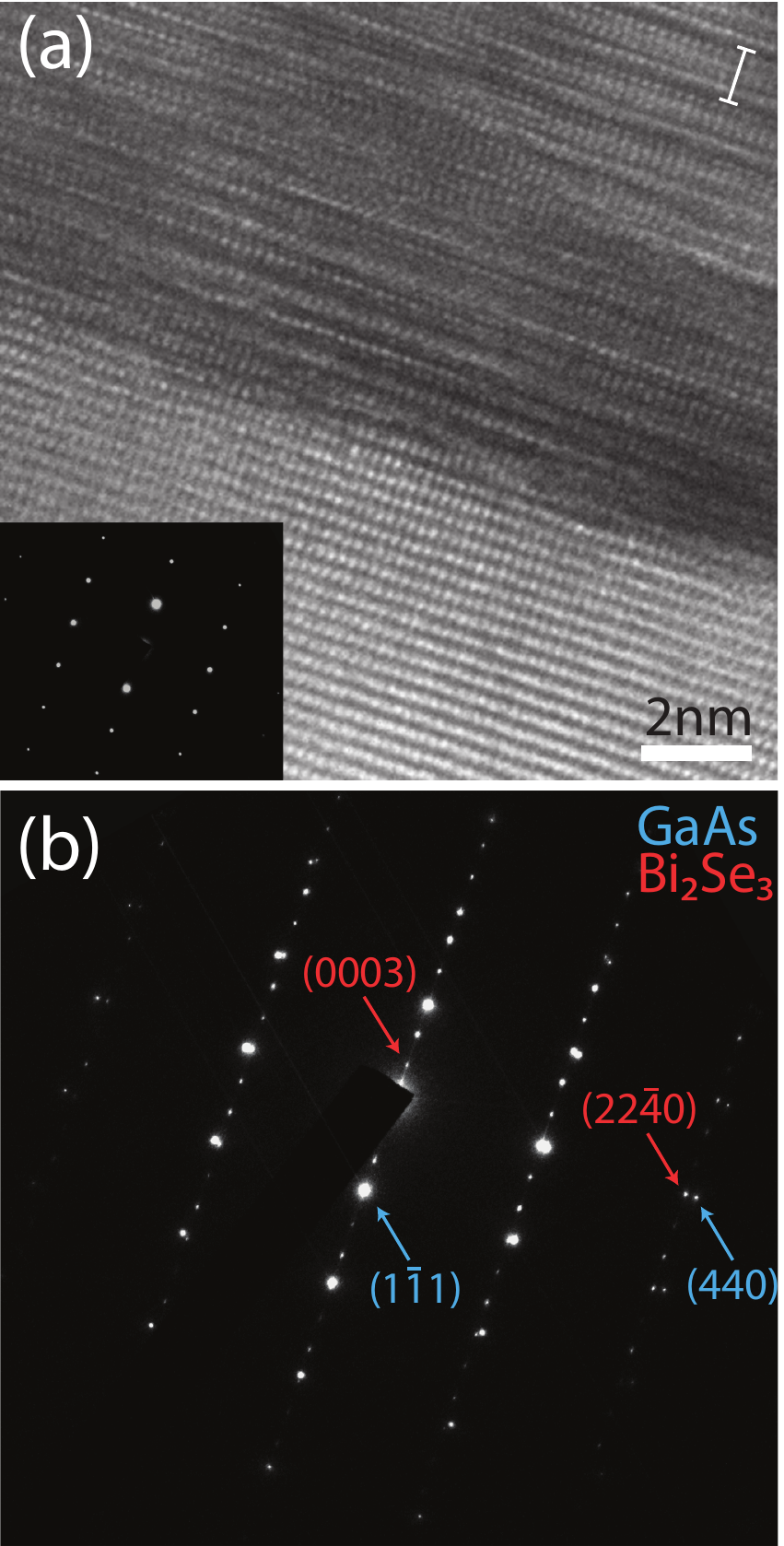}
\caption{\label{TEM}(Color online) (a) HRTEM image of the heterostructure showing epitaxial growth of Bi$_2$Se$_3$ on the GaAs substrate without the formation of an amorphous layer at the interface. The distance (0.98 nm) between QLs is shown at the top right. Inset shows the diffraction pattern of the substrate. (b) Diffraction from the whole area in (a) showing both the GaAs and Bi$_2$Se$_3$ patterns. The c-axis Bi$_2$Se$_3$ film grows in registry with the hexagonal GaAs 111B surface. The separation of the high index spots indicates that the film is relaxed in-plane.}
\end{figure}

To confirm the epitaxial growth of the Bi$_2$Se$_3$ thin film, we have carried out high-resolution transmission electron microscopy (HRTEM) on one of the samples grown directly on the thin GaAs buffer. Fig. ~\ref{TEM}(a) shows a typical HRTEM image at the interface of Bi$_2$Se$_3$ and GaAs. The lattice fringes of the phase contrast images show a good registry between the film and substrate without any amorphous growth or secondary phases occurring at the interface. The inset shows a selected area diffraction (SAD) pattern from just the GaAs substrate. Fig. ~\ref{TEM}(b) shows the SAD pattern from the whole region spanning the interface. Besides the pattern due to GaAs (blue indexes), the new spots (red indexes) are consistent with a single crystal Bi$_2$Se$_3$ film that has grown epitaxially on the GaAs. The interplanar distance between Bi$_2$Se$_3$ (22$\overline{4}$0) and GaAs (440) is found to be 0.336 nm$^{-1}$ in reciprocal space yielding a lattice mismatch in the $ab$ plane of 3.62\%, consistent with the expected value of 3.55\%, and indicating that the film is relaxed. Surprisingly, we do not find any evidence of twinning or dislocations in the TEM study, despite the large lattice mismatch. Both the HRTEM images and the diffraction patterns from several different areas show that the Bi$_2$Se$_3$ thin films are generally high-quality single crystals with a low density of defects. 

\begin{figure}
\includegraphics[width=50mm]{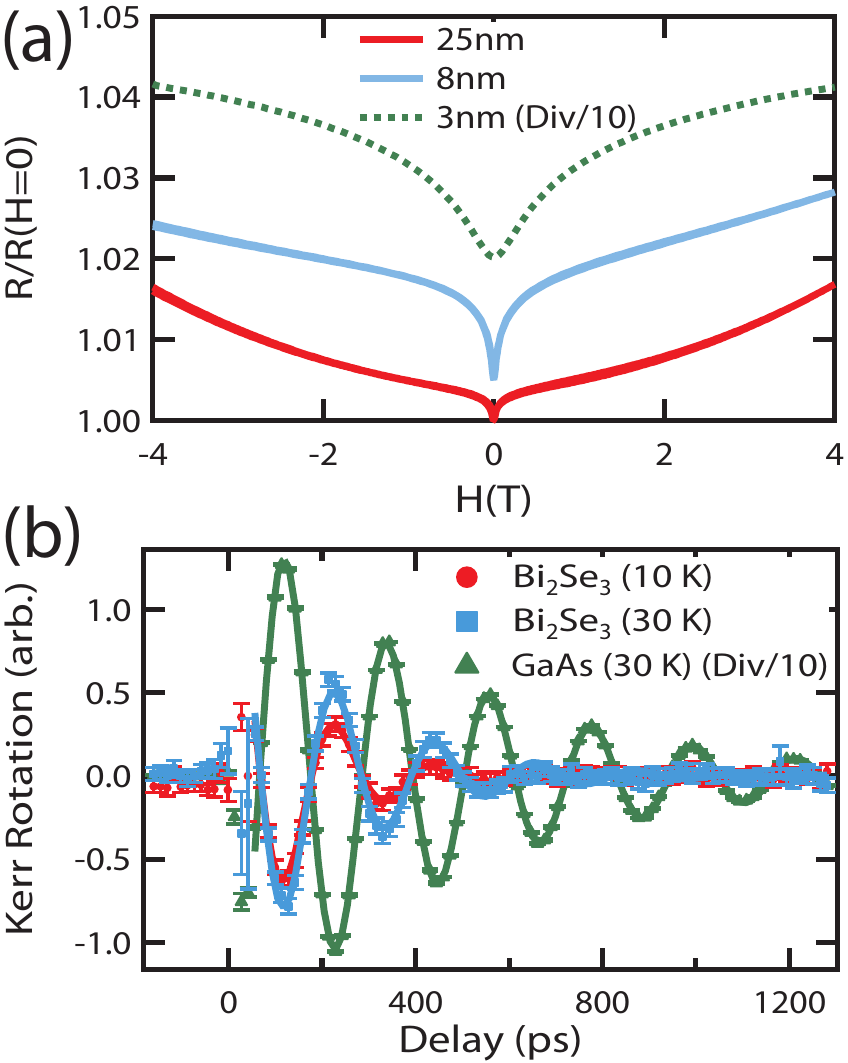}
\caption{\label{transport}(Color online) (a) Normalized MR at 4.2 K in three films with varying thickness ($t$) and carrier density ($n$): $t = 25$ nm, $n = 1.83 \times$10$^{19}$ cm$^{-3}$ (red); $t = 8$ nm, $n = 3.99 \times$10$^{19}$ cm$^{-3}$(blue); $t = 3$ nm, $n = 1.71 \times$10$^{19}$ cm$^{-3}$(dashed green). The normalized MR of the 3 nm film is divided by 10 and curves are offset for clarity. (b) Red circles and blue squares show the TRKR measured with an in-plane magnetic field of 0.75 T at the GaAs/Bi$_2$Se$_3$ interface in the 8 nm film described in (a). Green triangles show the TRKR (divided by 10) from the GaAs substrate alone. Pump and probe wavelength is 810 nm.}
\end{figure}

Electrical transport studies were carried out at 4.2 K using lithographically patterned and wet etched Hall bars (with dimensions of 650 $\mu$m $\times$ 400 $\mu$m) in perpendicular magnetic fields up to 4 T. Electrical and Hall conductivity measurements reveal that all the samples studied are n-doped with carrier densities in the range $8.06 \times10^{18}$ cm$^{-3} \lesssim n \lesssim 4 \times10^{19}$ cm$^{-3}$ and mobilities in the range $\sim 100 - \sim 1000$ cm$^{2}$ (V.s)$^{-1}$, consistent with previous reports of MBE growth.\cite{Li2010} Thus, we are still faced with unintentional background doping, presumably from a lack of stoichiometry and perhaps some contributions from unintentional Cd contamination from an earlier source in our MBE chamber. Magnetoresistance (MR) curves are shown in Fig.~\ref{transport}(a) for various film thicknesses. All show a positive MR cusp, consistent with weak anti-localization corrections to diffusive transport and typical of measurements of Bi$_2$Se$_3$ reported in the literature.\cite{Checkelsky2009,Peng2010} A systematic analysis of the temperature, magnetic field and sample thickness dependence of the MR will be reported elsewhere. 

Finally, we discuss preliminary magneto-optical measurements that probe spin-dependent phenomena associated with the interface in these heterostructures. We used a well-established  time-resolved Kerr rotation (TRKR) technique \cite{PhysRevLett.80.4313} to demonstrate a possible method of probing spin polarization in a TI via coupling to spin states in a conventional semiconductor. Figure 4(b) shows TRKR curves for optically-injected spins in the GaAs substrate precessing in an in-plane magnetic field. Data measured through an 8 nm layer of Bi$_2$Se$_3$ are shown at two temperatures, along with reference data from an area where the Bi$_2$Se$_3$ layer was wet-etched away. By fitting the TRKR to a damped sinusoid,\cite{PhysRevLett.80.4313} we deduce the g-factor and the inhomogeneous spin lifetime (${T_2}^\ast$). While the g-factor of spins in GaAs ($g = -0.44$) is unchanged by overgrowth of Bi$_2$Se$_3$, ${T_2}^\ast$ is significantly shorter at the Bi$_2$Se$_3$ interface: at $T = 30$ K, ${T_2}^\ast = 160$ ps at the interface, compared with ${T_2}^\ast = 450$ ps in the reference region.

In summary, we have demonstrated the coherent epitaxial growth of the candidate TI material Bi$_2$Se$_3$ on GaAs (111)B substrates. The ability to synthesize Bi$_2$Se$_3$ epitaxial films with high quality heterointerfaces on GaAs and ZnSe opens the door to a host of interesting heterostructure applications, including TI-magnetic semiconductor interfaces, where magnetic monopoles or Majorana fermions at domain walls could be studied.\cite{Qi2009, Akhmerov2009} 

We thank Josh Maier for preparing the cross-sectional TEM specimen. This work was supported by the Penn State MRSEC (NSF-DMR-0820404) through a seed grant and the REU program, and partially by grants ONR N00014-09-1-0221 and -0309. We acknowledge use of the NSF National Nanofabrication Users Network Facilities at Penn State and UCSB.


\begin{thebibliography}{19}%
\makeatletter
\providecommand \@ifxundefined [1]{%
 \@ifx{#1\undefined}
}%
\providecommand \@ifnum [1]{%
 \ifnum #1\expandafter \@firstoftwo
 \else \expandafter \@secondoftwo
 \fi
}%
\providecommand \@ifx [1]{%
 \ifx #1\expandafter \@firstoftwo
 \else \expandafter \@secondoftwo
 \fi
}%
\providecommand \natexlab [1]{#1}%
\providecommand \enquote  [1]{``#1''}%
\providecommand \bibnamefont  [1]{#1}%
\providecommand \bibfnamefont [1]{#1}%
\providecommand \citenamefont [1]{#1}%
\providecommand \href@noop [0]{\@secondoftwo}%
\providecommand \href [0]{\begingroup \@sanitize@url \@href}%
\providecommand \@href[1]{\@@startlink{#1}\@@href}%
\providecommand \@@href[1]{\endgroup#1\@@endlink}%
\providecommand \@sanitize@url [0]{\catcode `\\12\catcode `\$12\catcode
  `\&12\catcode `\#12\catcode `\^12\catcode `\_12\catcode `\%12\relax}%
\providecommand \@@startlink[1]{}%
\providecommand \@@endlink[0]{}%
\providecommand \url  [0]{\begingroup\@sanitize@url \@url }%
\providecommand \@url [1]{\endgroup\@href {#1}{\urlprefix }}%
\providecommand \urlprefix  [0]{URL }%
\providecommand \Eprint [0]{\href }%
\providecommand \doibase [0]{http://dx.doi.org/}%
\providecommand \selectlanguage [0]{\@gobble}%
\providecommand \bibinfo  [0]{\@secondoftwo}%
\providecommand \bibfield  [0]{\@secondoftwo}%
\providecommand \translation [1]{[#1]}%
\providecommand \BibitemOpen [0]{}%
\providecommand \bibitemStop [0]{}%
\providecommand \bibitemNoStop [0]{.\EOS\space}%
\providecommand \EOS [0]{\spacefactor3000\relax}%
\providecommand \BibitemShut  [1]{\csname bibitem#1\endcsname}%
\let\auto@bib@innerbib\@empty
\bibitem [{\citenamefont {Fu}, \citenamefont {Kane},\ and\ \citenamefont
  {Mele}(2007)}]{Fu:2007fk}%
  \BibitemOpen
  \bibfield  {author} {\bibinfo {author} {\bibfnamefont {L.}~\bibnamefont
  {Fu}}, \bibinfo {author} {\bibfnamefont {C.~L.}\ \bibnamefont {Kane}}, \ and\
  \bibinfo {author} {\bibfnamefont {E.~J.}\ \bibnamefont {Mele}},\ }\href
  {\doibase 10.1103/PhysRevLett.98.106803} {\bibfield  {journal} {\bibinfo
  {journal} {Phys. Rev. Lett.}\ }\textbf {\bibinfo {volume} {98}},\ \bibinfo
  {pages} {106803} (\bibinfo {year} {2007})}\BibitemShut {NoStop}%
\bibitem [{\citenamefont {Qi}, \citenamefont {Hughes},\ and\ \citenamefont
  {Zhang}(2008)}]{Qi2008}%
  \BibitemOpen
  \bibfield  {author} {\bibinfo {author} {\bibfnamefont {X.-L.}\ \bibnamefont
  {Qi}}, \bibinfo {author} {\bibfnamefont {T.~L.}\ \bibnamefont {Hughes}}, \
  and\ \bibinfo {author} {\bibfnamefont {S.-C.}\ \bibnamefont {Zhang}},\ }\href
  {\doibase 10.1103/PhysRevB.78.195424} {\bibfield  {journal} {\bibinfo
  {journal} {Phys. Rev. B}\ }\textbf {\bibinfo {volume} {78}},\ \bibinfo
  {pages} {195424} (\bibinfo {year} {2008})}\BibitemShut {NoStop}%
\bibitem [{\citenamefont {Zhang}\ \emph
  {et~al.}(2009{\natexlab{a}})\citenamefont {Zhang}, \citenamefont {Liu},
  \citenamefont {Qi}, \citenamefont {Dai}, \citenamefont {Fang},\ and\
  \citenamefont {Zhang}}]{Zhang2009a}%
  \BibitemOpen
  \bibfield  {author} {\bibinfo {author} {\bibfnamefont {H.}~\bibnamefont
  {Zhang}}, \bibinfo {author} {\bibfnamefont {C.-X.}\ \bibnamefont {Liu}},
  \bibinfo {author} {\bibfnamefont {X.-L.}\ \bibnamefont {Qi}}, \bibinfo
  {author} {\bibfnamefont {X.}~\bibnamefont {Dai}}, \bibinfo {author}
  {\bibfnamefont {Z.}~\bibnamefont {Fang}}, \ and\ \bibinfo {author}
  {\bibfnamefont {S.-C.}\ \bibnamefont {Zhang}},\ }\href
  {http://dx.doi.org/10.1038/nphys1270} {\bibfield  {journal} {\bibinfo
  {journal} {Nat. Phys.}\ }\textbf {\bibinfo {volume} {5}},\ \bibinfo {pages}
  {438} (\bibinfo {year} {2009}{\natexlab{a}})}\BibitemShut {NoStop}%
\bibitem [{\citenamefont {Xia}\ \emph {et~al.}(2009)\citenamefont {Xia},
  \citenamefont {Qian}, \citenamefont {Hsieh}, \citenamefont {Wray},
  \citenamefont {Pal}, \citenamefont {Lin}, \citenamefont {Bansil},
  \citenamefont {Grauer}, \citenamefont {Hor}, \citenamefont {Cava},\ and\
  \citenamefont {Hasan}}]{Xia2009}%
  \BibitemOpen
  \bibfield  {author} {\bibinfo {author} {\bibfnamefont {Y.}~\bibnamefont
  {Xia}}, \bibinfo {author} {\bibfnamefont {D.}~\bibnamefont {Qian}}, \bibinfo
  {author} {\bibfnamefont {D.}~\bibnamefont {Hsieh}}, \bibinfo {author}
  {\bibfnamefont {L.}~\bibnamefont {Wray}}, \bibinfo {author} {\bibfnamefont
  {A.}~\bibnamefont {Pal}}, \bibinfo {author} {\bibfnamefont {H.}~\bibnamefont
  {Lin}}, \bibinfo {author} {\bibfnamefont {A.}~\bibnamefont {Bansil}},
  \bibinfo {author} {\bibfnamefont {D.}~\bibnamefont {Grauer}}, \bibinfo
  {author} {\bibfnamefont {Y.~S.}\ \bibnamefont {Hor}}, \bibinfo {author}
  {\bibfnamefont {R.~J.}\ \bibnamefont {Cava}}, \ and\ \bibinfo {author}
  {\bibfnamefont {M.~Z.}\ \bibnamefont {Hasan}},\ }\href
  {http://dx.doi.org/10.1038/nphys1274} {\bibfield  {journal} {\bibinfo
  {journal} {Nat. Phys.}\ }\textbf {\bibinfo {volume} {5}},\ \bibinfo {pages}
  {398} (\bibinfo {year} {2009})}\BibitemShut {NoStop}%
\bibitem [{\citenamefont {Hsieh}\ \emph {et~al.}(2009)\citenamefont {Hsieh},
  \citenamefont {Xia}, \citenamefont {Qian}, \citenamefont {Wray},
  \citenamefont {Dil}, \citenamefont {Meier}, \citenamefont {Osterwalder},
  \citenamefont {Patthey}, \citenamefont {Checkelsky}, \citenamefont {Ong},
  \citenamefont {Fedorov}, \citenamefont {Lin}, \citenamefont {Bansil},
  \citenamefont {Grauer}, \citenamefont {Hor}, \citenamefont {Cava},\ and\
  \citenamefont {Hasan}}]{Hsieh2009}%
  \BibitemOpen
  \bibfield  {author} {\bibinfo {author} {\bibfnamefont {D.}~\bibnamefont
  {Hsieh}}, \bibinfo {author} {\bibfnamefont {Y.}~\bibnamefont {Xia}}, \bibinfo
  {author} {\bibfnamefont {D.}~\bibnamefont {Qian}}, \bibinfo {author}
  {\bibfnamefont {L.}~\bibnamefont {Wray}}, \bibinfo {author} {\bibfnamefont
  {J.~H.}\ \bibnamefont {Dil}}, \bibinfo {author} {\bibfnamefont
  {F.}~\bibnamefont {Meier}}, \bibinfo {author} {\bibfnamefont
  {J.}~\bibnamefont {Osterwalder}}, \bibinfo {author} {\bibfnamefont
  {L.}~\bibnamefont {Patthey}}, \bibinfo {author} {\bibfnamefont {J.~G.}\
  \bibnamefont {Checkelsky}}, \bibinfo {author} {\bibfnamefont {N.~P.}\
  \bibnamefont {Ong}}, \bibinfo {author} {\bibfnamefont {A.~V.}\ \bibnamefont
  {Fedorov}}, \bibinfo {author} {\bibfnamefont {H.}~\bibnamefont {Lin}},
  \bibinfo {author} {\bibfnamefont {A.}~\bibnamefont {Bansil}}, \bibinfo
  {author} {\bibfnamefont {D.}~\bibnamefont {Grauer}}, \bibinfo {author}
  {\bibfnamefont {Y.~S.}\ \bibnamefont {Hor}}, \bibinfo {author} {\bibfnamefont
  {R.~J.}\ \bibnamefont {Cava}}, \ and\ \bibinfo {author} {\bibfnamefont
  {M.~Z.}\ \bibnamefont {Hasan}},\ }\href
  {http://dx.doi.org/10.1038/nature08234} {\bibfield  {journal} {\bibinfo
  {journal} {Nature}\ }\textbf {\bibinfo {volume} {460}},\ \bibinfo {pages}
  {1101} (\bibinfo {year} {2009})}\BibitemShut {NoStop}%
\bibitem [{\citenamefont {Navr\'atil}\ \emph {et~al.}(2004)\citenamefont
  {Navr\'atil}, \citenamefont {Hor\a'k}, \citenamefont {Plechcek}, \citenamefont
  {Kamba}, \citenamefont {Lostak}, \citenamefont {Dyck}, \citenamefont {Chen},\
  and\ \citenamefont {Uher}}]{Navratil2004}%
  \BibitemOpen
  \bibfield  {author} {\bibinfo {author} {\bibfnamefont {J.}~\bibnamefont
  {Navr\'atil}}, \bibinfo {author} {\bibfnamefont {J.}~\bibnamefont {Hor\'ak}},
  \bibinfo {author} {\bibfnamefont {T.}~\bibnamefont {Plechacek}}, \bibinfo
  {author} {\bibfnamefont {S.}~\bibnamefont {Kamba}}, \bibinfo {author}
  {\bibfnamefont {P.}~\bibnamefont {Lostak}}, \bibinfo {author} {\bibfnamefont
  {J.}~\bibnamefont {Dyck}}, \bibinfo {author} {\bibfnamefont {W.}~\bibnamefont
  {Chen}}, \ and\ \bibinfo {author} {\bibfnamefont {C.}~\bibnamefont {Uher}},\
  }\href {\doibase DOI: 10.1016/j.jssc.2003.12.031} {\bibfield  {journal}
  {\bibinfo  {journal} {J. Solid State Chem.}\ }\textbf {\bibinfo {volume}
  {177}},\ \bibinfo {pages} {1704 } (\bibinfo {year} {2004})}\BibitemShut
  {NoStop}%
\bibitem [{\citenamefont {Hor}\ \emph {et~al.}(2009)\citenamefont {Hor},
  \citenamefont {Richardella}, \citenamefont {Roushan}, \citenamefont {Xia},
  \citenamefont {Checkelsky}, \citenamefont {Yazdani}, \citenamefont {Hasan},
  \citenamefont {Ong},\ and\ \citenamefont {Cava}}]{Hor2009}%
  \BibitemOpen
  \bibfield  {author} {\bibinfo {author} {\bibfnamefont {Y.~S.}\ \bibnamefont
  {Hor}}, \bibinfo {author} {\bibfnamefont {A.}~\bibnamefont {Richardella}},
  \bibinfo {author} {\bibfnamefont {P.}~\bibnamefont {Roushan}}, \bibinfo
  {author} {\bibfnamefont {Y.}~\bibnamefont {Xia}}, \bibinfo {author}
  {\bibfnamefont {J.~G.}\ \bibnamefont {Checkelsky}}, \bibinfo {author}
  {\bibfnamefont {A.}~\bibnamefont {Yazdani}}, \bibinfo {author} {\bibfnamefont
  {M.~Z.}\ \bibnamefont {Hasan}}, \bibinfo {author} {\bibfnamefont {N.~P.}\
  \bibnamefont {Ong}}, \ and\ \bibinfo {author} {\bibfnamefont {R.~J.}\
  \bibnamefont {Cava}},\ }\href {\doibase 10.1103/PhysRevB.79.195208}
  {\bibfield  {journal} {\bibinfo  {journal} {Phys. Rev. B}\ }\textbf {\bibinfo
  {volume} {79}},\ \bibinfo {pages} {195208} (\bibinfo {year}
  {2009})}\BibitemShut {NoStop}%
\bibitem [{\citenamefont {Zhang}\ \emph
  {et~al.}(2009{\natexlab{b}})\citenamefont {Zhang}, \citenamefont {Qin},
  \citenamefont {Teng}, \citenamefont {Guo}, \citenamefont {Guo}, \citenamefont
  {Dai}, \citenamefont {Fang},\ and\ \citenamefont {Wu}}]{Zhang2009}%
  \BibitemOpen
  \bibfield  {author} {\bibinfo {author} {\bibfnamefont {G.}~\bibnamefont
  {Zhang}}, \bibinfo {author} {\bibfnamefont {H.}~\bibnamefont {Qin}}, \bibinfo
  {author} {\bibfnamefont {J.}~\bibnamefont {Teng}}, \bibinfo {author}
  {\bibfnamefont {J.}~\bibnamefont {Guo}}, \bibinfo {author} {\bibfnamefont
  {Q.}~\bibnamefont {Guo}}, \bibinfo {author} {\bibfnamefont {X.}~\bibnamefont
  {Dai}}, \bibinfo {author} {\bibfnamefont {Z.}~\bibnamefont {Fang}}, \ and\
  \bibinfo {author} {\bibfnamefont {K.}~\bibnamefont {Wu}},\ }\href {\doibase
  10.1063/1.3200237} {\bibfield  {journal} {\bibinfo  {journal} {Appl. Phys.
  Lett.}\ }\textbf {\bibinfo {volume} {95}},\ \bibinfo {eid} {053114} (\bibinfo
  {year} {2009}{\natexlab{b}})}\BibitemShut {NoStop}%
\bibitem [{\citenamefont {{Li}}\ \emph {et~al.}(2010)\citenamefont {{Li}},
  \citenamefont {{Wang}}, \citenamefont {{Kan}}, \citenamefont {{Guo}},
  \citenamefont {{He}}, \citenamefont {{Wang}}, \citenamefont {{Wang}},
  \citenamefont {{Wong}}, \citenamefont {{Wang}},\ and\ \citenamefont
  {{Xie}}}]{Li2010}%
  \BibitemOpen
  \bibfield  {author} {\bibinfo {author} {\bibfnamefont {H.~D.}\ \bibnamefont
  {{Li}}}, \bibinfo {author} {\bibfnamefont {Z.~Y.}\ \bibnamefont {{Wang}}},
  \bibinfo {author} {\bibfnamefont {X.}~\bibnamefont {{Kan}}}, \bibinfo
  {author} {\bibfnamefont {X.}~\bibnamefont {{Guo}}}, \bibinfo {author}
  {\bibfnamefont {H.~T.}\ \bibnamefont {{He}}}, \bibinfo {author}
  {\bibfnamefont {Z.}~\bibnamefont {{Wang}}}, \bibinfo {author} {\bibfnamefont
  {J.~N.}\ \bibnamefont {{Wang}}}, \bibinfo {author} {\bibfnamefont {T.~L.}\
  \bibnamefont {{Wong}}}, \bibinfo {author} {\bibfnamefont {N.}~\bibnamefont
  {{Wang}}}, \ and\ \bibinfo {author} {\bibfnamefont {M.~H.}\ \bibnamefont
  {{Xie}}},\ }\href@noop {} {\bibfield  {journal} {\bibinfo  {journal} {ArXiv
  e-prints}\ } (\bibinfo {year} {2010})},\ \Eprint
  {http://arxiv.org/abs/1005.0449} {1005.0449} \BibitemShut {NoStop}%
\bibitem [{\citenamefont {Zhang}\ \emph {et~al.}(2010)\citenamefont {Zhang},
  \citenamefont {He}, \citenamefont {Chang}, \citenamefont {Song},
  \citenamefont {Wang}, \citenamefont {Chen}, \citenamefont {Jia},
  \citenamefont {Fang}, \citenamefont {Dai}, \citenamefont {Shan},
  \citenamefont {Shen}, \citenamefont {Niu}, \citenamefont {Qi}, \citenamefont
  {Zhang}, \citenamefont {Ma},\ and\ \citenamefont {Xue}}]{Zhang2010}%
  \BibitemOpen
  \bibfield  {author} {\bibinfo {author} {\bibfnamefont {Y.}~\bibnamefont
  {Zhang}}, \bibinfo {author} {\bibfnamefont {K.}~\bibnamefont {He}}, \bibinfo
  {author} {\bibfnamefont {C.-Z.}\ \bibnamefont {Chang}}, \bibinfo {author}
  {\bibfnamefont {C.-L.}\ \bibnamefont {Song}}, \bibinfo {author}
  {\bibfnamefont {L.-L.}\ \bibnamefont {Wang}}, \bibinfo {author}
  {\bibfnamefont {X.}~\bibnamefont {Chen}}, \bibinfo {author} {\bibfnamefont
  {J.-F.}\ \bibnamefont {Jia}}, \bibinfo {author} {\bibfnamefont
  {Z.}~\bibnamefont {Fang}}, \bibinfo {author} {\bibfnamefont {X.}~\bibnamefont
  {Dai}}, \bibinfo {author} {\bibfnamefont {W.-Y.}\ \bibnamefont {Shan}},
  \bibinfo {author} {\bibfnamefont {S.-Q.}\ \bibnamefont {Shen}}, \bibinfo
  {author} {\bibfnamefont {Q.}~\bibnamefont {Niu}}, \bibinfo {author}
  {\bibfnamefont {X.-L.}\ \bibnamefont {Qi}}, \bibinfo {author} {\bibfnamefont
  {S.-C.}\ \bibnamefont {Zhang}}, \bibinfo {author} {\bibfnamefont {X.-C.}\
  \bibnamefont {Ma}}, \ and\ \bibinfo {author} {\bibfnamefont {Q.-K.}\
  \bibnamefont {Xue}},\ }\href {http://dx.doi.org/10.1038/nphys1689} {\bibfield
   {journal} {\bibinfo  {journal} {Nat. Phys.}\ }\textbf {\bibinfo {volume}
  {6}},\ \bibinfo {pages} {584} (\bibinfo {year} {2010})}\BibitemShut {NoStop}%
\bibitem [{\citenamefont {Song}\ \emph {et~al.}(2010)\citenamefont {Song},
  \citenamefont {Wang}, \citenamefont {Jiang}, \citenamefont {Zhang},
  \citenamefont {Chang}, \citenamefont {Wang}, \citenamefont {He},
  \citenamefont {Chen}, \citenamefont {Jia}, \citenamefont {Wang},
  \citenamefont {Fang}, \citenamefont {Dai}, \citenamefont {Xie}, \citenamefont
  {Qi}, \citenamefont {Zhang}, \citenamefont {Xue},\ and\ \citenamefont
  {Ma}}]{Song2010}%
  \BibitemOpen
  \bibfield  {author} {\bibinfo {author} {\bibfnamefont {C.-L.}\ \bibnamefont
  {Song}}, \bibinfo {author} {\bibfnamefont {Y.-L.}\ \bibnamefont {Wang}},
  \bibinfo {author} {\bibfnamefont {Y.-P.}\ \bibnamefont {Jiang}}, \bibinfo
  {author} {\bibfnamefont {Y.}~\bibnamefont {Zhang}}, \bibinfo {author}
  {\bibfnamefont {C.-Z.}\ \bibnamefont {Chang}}, \bibinfo {author}
  {\bibfnamefont {L.}~\bibnamefont {Wang}}, \bibinfo {author} {\bibfnamefont
  {K.}~\bibnamefont {He}}, \bibinfo {author} {\bibfnamefont {X.}~\bibnamefont
  {Chen}}, \bibinfo {author} {\bibfnamefont {J.-F.}\ \bibnamefont {Jia}},
  \bibinfo {author} {\bibfnamefont {Y.}~\bibnamefont {Wang}}, \bibinfo {author}
  {\bibfnamefont {Z.}~\bibnamefont {Fang}}, \bibinfo {author} {\bibfnamefont
  {X.}~\bibnamefont {Dai}}, \bibinfo {author} {\bibfnamefont {X.-C.}\
  \bibnamefont {Xie}}, \bibinfo {author} {\bibfnamefont {X.-L.}\ \bibnamefont
  {Qi}}, \bibinfo {author} {\bibfnamefont {S.-C.}\ \bibnamefont {Zhang}},
  \bibinfo {author} {\bibfnamefont {Q.-K.}\ \bibnamefont {Xue}}, \ and\
  \bibinfo {author} {\bibfnamefont {X.}~\bibnamefont {Ma}},\ }\href {\doibase
  10.1063/1.3494595} {\bibfield  {journal} {\bibinfo  {journal} {Appl. Phys.
  Lett.}\ }\textbf {\bibinfo {volume} {97}},\ \bibinfo {eid} {143118} (\bibinfo
  {year} {2010})}\BibitemShut {NoStop}%
\bibitem [{\citenamefont {Chen}\ \emph {et~al.}(2010)\citenamefont {Chen},
  \citenamefont {Qin}, \citenamefont {Yang}, \citenamefont {Liu}, \citenamefont
  {Guan}, \citenamefont {Qu}, \citenamefont {Zhang}, \citenamefont {Shi},
  \citenamefont {Xie}, \citenamefont {Yang}, \citenamefont {Wu}, \citenamefont
  {Li},\ and\ \citenamefont {Lu}}]{Chen2010}%
  \BibitemOpen
  \bibfield  {author} {\bibinfo {author} {\bibfnamefont {J.}~\bibnamefont
  {Chen}}, \bibinfo {author} {\bibfnamefont {H.~J.}\ \bibnamefont {Qin}},
  \bibinfo {author} {\bibfnamefont {F.}~\bibnamefont {Yang}}, \bibinfo {author}
  {\bibfnamefont {J.}~\bibnamefont {Liu}}, \bibinfo {author} {\bibfnamefont
  {T.}~\bibnamefont {Guan}}, \bibinfo {author} {\bibfnamefont {F.~M.}\
  \bibnamefont {Qu}}, \bibinfo {author} {\bibfnamefont {G.~H.}\ \bibnamefont
  {Zhang}}, \bibinfo {author} {\bibfnamefont {J.~R.}\ \bibnamefont {Shi}},
  \bibinfo {author} {\bibfnamefont {X.~C.}\ \bibnamefont {Xie}}, \bibinfo
  {author} {\bibfnamefont {C.~L.}\ \bibnamefont {Yang}}, \bibinfo {author}
  {\bibfnamefont {K.~H.}\ \bibnamefont {Wu}}, \bibinfo {author} {\bibfnamefont
  {Y.~Q.}\ \bibnamefont {Li}}, \ and\ \bibinfo {author} {\bibfnamefont
  {L.}~\bibnamefont {Lu}},\ }\href {\doibase 10.1103/PhysRevLett.105.176602}
  {\bibfield  {journal} {\bibinfo  {journal} {Phys. Rev. Lett.}\ }\textbf
  {\bibinfo {volume} {105}},\ \bibinfo {pages} {176602} (\bibinfo {year}
  {2010})}\BibitemShut {NoStop}%
\bibitem [{\citenamefont {Sugahara}\ and\ \citenamefont
  {Tanaka}(2003)}]{Sugahara2003}%
  \BibitemOpen
  \bibfield  {author} {\bibinfo {author} {\bibfnamefont {S.}~\bibnamefont
  {Sugahara}}\ and\ \bibinfo {author} {\bibfnamefont {M.}~\bibnamefont
  {Tanaka}},\ }\href {\doibase DOI: 10.1016/S0022-0248(02)02280-7} {\bibfield
  {journal} {\bibinfo  {journal} {J. Cryst. Growth}\ }\textbf {\bibinfo
  {volume} {251}},\ \bibinfo {pages} {317 } (\bibinfo {year}
  {2003})}\BibitemShut {NoStop}%
\bibitem [{\citenamefont {Checkelsky}\ \emph {et~al.}(2009)\citenamefont
  {Checkelsky}, \citenamefont {Hor}, \citenamefont {Liu}, \citenamefont {Qu},
  \citenamefont {Cava},\ and\ \citenamefont {Ong}}]{Checkelsky2009}%
  \BibitemOpen
  \bibfield  {author} {\bibinfo {author} {\bibfnamefont {J.~G.}\ \bibnamefont
  {Checkelsky}}, \bibinfo {author} {\bibfnamefont {Y.~S.}\ \bibnamefont {Hor}},
  \bibinfo {author} {\bibfnamefont {M.-H.}\ \bibnamefont {Liu}}, \bibinfo
  {author} {\bibfnamefont {D.-X.}\ \bibnamefont {Qu}}, \bibinfo {author}
  {\bibfnamefont {R.~J.}\ \bibnamefont {Cava}}, \ and\ \bibinfo {author}
  {\bibfnamefont {N.~P.}\ \bibnamefont {Ong}},\ }\href {\doibase
  10.1103/PhysRevLett.103.246601} {\bibfield  {journal} {\bibinfo  {journal}
  {Phys. Rev. Lett.}\ }\textbf {\bibinfo {volume} {103}},\ \bibinfo {pages}
  {246601} (\bibinfo {year} {2009})}\BibitemShut {NoStop}%
\bibitem [{\citenamefont {Peng}\ \emph {et~al.}(2010)\citenamefont {Peng},
  \citenamefont {Lai}, \citenamefont {Kong}, \citenamefont {Meister},
  \citenamefont {Chen}, \citenamefont {Qi}, \citenamefont {Zhang},
  \citenamefont {Shen},\ and\ \citenamefont {Cui}}]{Peng2010}%
  \BibitemOpen
  \bibfield  {author} {\bibinfo {author} {\bibfnamefont {H.}~\bibnamefont
  {Peng}}, \bibinfo {author} {\bibfnamefont {K.}~\bibnamefont {Lai}}, \bibinfo
  {author} {\bibfnamefont {D.}~\bibnamefont {Kong}}, \bibinfo {author}
  {\bibfnamefont {S.}~\bibnamefont {Meister}}, \bibinfo {author} {\bibfnamefont
  {Y.}~\bibnamefont {Chen}}, \bibinfo {author} {\bibfnamefont {X.-L.}\
  \bibnamefont {Qi}}, \bibinfo {author} {\bibfnamefont {S.-C.}\ \bibnamefont
  {Zhang}}, \bibinfo {author} {\bibfnamefont {Z.-X.}\ \bibnamefont {Shen}}, \
  and\ \bibinfo {author} {\bibfnamefont {Y.}~\bibnamefont {Cui}},\ }\href
  {http://dx.doi.org/10.1038/nmat2609} {\bibfield  {journal} {\bibinfo
  {journal} {Nat Mater}\ }\textbf {\bibinfo {volume} {9}},\ \bibinfo {pages}
  {225} (\bibinfo {year} {2010})}\BibitemShut {NoStop}%
\bibitem [{\citenamefont {Kikkawa}\ and\ \citenamefont
  {Awschalom}(1998)}]{PhysRevLett.80.4313}%
  \BibitemOpen
  \bibfield  {author} {\bibinfo {author} {\bibfnamefont {J.~M.}\ \bibnamefont
  {Kikkawa}}\ and\ \bibinfo {author} {\bibfnamefont {D.~D.}\ \bibnamefont
  {Awschalom}},\ }\href {\doibase 10.1103/PhysRevLett.80.4313} {\bibfield
  {journal} {\bibinfo  {journal} {Phys. Rev. Lett.}\ }\textbf {\bibinfo
  {volume} {80}},\ \bibinfo {pages} {4313} (\bibinfo {year}
  {1998})}\BibitemShut {NoStop}%
\bibitem [{\citenamefont {Qi}\ \emph {et~al.}(2009)\citenamefont {Qi},
  \citenamefont {Li}, \citenamefont {Zang},\ and\ \citenamefont
  {Zhang}}]{Qi2009}%
  \BibitemOpen
  \bibfield  {author} {\bibinfo {author} {\bibfnamefont {X.-L.}\ \bibnamefont
  {Qi}}, \bibinfo {author} {\bibfnamefont {R.}~\bibnamefont {Li}}, \bibinfo
  {author} {\bibfnamefont {J.}~\bibnamefont {Zang}}, \ and\ \bibinfo {author}
  {\bibfnamefont {S.-C.}\ \bibnamefont {Zhang}},\ }\href {\doibase
  10.1126/science.1167747} {\bibfield  {journal} {\bibinfo  {journal}
  {Science}\ }\textbf {\bibinfo {volume} {323}},\ \bibinfo {pages} {1184}
  (\bibinfo {year} {2009})}\BibitemShut {NoStop}%
\bibitem [{\citenamefont {Akhmerov}, \citenamefont {Nilsson},\ and\
  \citenamefont {Beenakker}(2009)}]{Akhmerov2009}%
  \BibitemOpen
  \bibfield  {author} {\bibinfo {author} {\bibfnamefont {A.~R.}\ \bibnamefont
  {Akhmerov}}, \bibinfo {author} {\bibfnamefont {J.}~\bibnamefont {Nilsson}}, \
  and\ \bibinfo {author} {\bibfnamefont {C.~W.~J.}\ \bibnamefont {Beenakker}},\
  }\href {\doibase 10.1103/PhysRevLett.102.216404} {\bibfield  {journal}
  {\bibinfo  {journal} {Phys. Rev. Lett.}\ }\textbf {\bibinfo {volume} {102}},\
  \bibinfo {pages} {216404} (\bibinfo {year} {2009})}\BibitemShut {NoStop}%
\end{thebibliography}
%

\end{document}